\documentclass[conference,compsoc]{IEEEtran}
\IEEEoverridecommandlockouts
\usepackage{cite}
\usepackage{amsmath,amssymb,amsfonts}
\usepackage{algorithm2e}
\usepackage[most]{tcolorbox}

\newtcolorbox{summarybox}[2][]{
    lower separated=false,         %
    colback=white,                 %
    colframe=black,                %
    fonttitle=\bfseries,           %
    colbacktitle=black,            %
    coltitle=white,                %
    enhanced,                      %
    attach boxed title to top left={yshift=-0.1in, xshift=0.15in}, %
    boxed title style={boxrule=0pt, colframe=white}, %
    title=#2,                      %
    #1                             %
}

\RestyleAlgo{ruled}  %
\SetKwComment{Comment}{// }{}  %

\usepackage{balance}
\usepackage{enumitem}
\usepackage{graphicx}
\usepackage{float}
\usepackage{textcomp}
\usepackage{xcolor}
\usepackage{hyperref}
\usepackage{caption,subcaption}
\usepackage{pifont} %
\usepackage{array} %
\usepackage{tabularx} %
\usepackage{xurl}
\usepackage{hyperref}
\usepackage{multicol}
\usepackage[capitalise,noabbrev]{cleveref} 
\crefname{appsec}{Appendix}{Appendices} 

\usepackage{siunitx}
\def\BibTeX{{\rm B\kern-.05em{\sc i\kern-.025em b}\kern-.08em
    T\kern-.1667em\lower.7ex\hbox{E}\kern-.125emX}}
\begin{document}

\newcommand{\anindya}[1]{{\color{red}[AM: #1]}}
\newcommand{\marc}[1]{{\color{magenta}[MB: #1]}}
\newcommand{\rw}[1]{{\color{green}[RW: #1]}}
\newcommand{\scott}[1]{{\color{brown}[SS: #1]}}
\newcommand{\mj}[1]{{\color{blue}[MJ: #1]}}

\hyphenation{NinjaDoH}
\newcommand{\name}{\textit{NinjaDoH}}

\title{NinjaDoH: A Censorship-Resistant Moving Target DoH Server \\Using Hyperscalers and IPNS}

\author{
    \IEEEauthorblockN{
        Scott Seidenberger\IEEEauthorrefmark{1},
        Marc Beret\IEEEauthorrefmark{1},
        Raveen Wijewickrama\IEEEauthorrefmark{2},
        Murtuza Jadliwala\IEEEauthorrefmark{2}, and
        Anindya Maiti\IEEEauthorrefmark{1}
    }
    \IEEEauthorblockA{\IEEEauthorrefmark{1}University of Oklahoma, Norman, OK, USA\\
    Email: \{seidenberger, marc.beret, am\}@ou.edu}
    \IEEEauthorblockA{\IEEEauthorrefmark{2}University of Texas at San Antonio, San Antonio, TX, USA\\
    Email: \{raveen.wijewickrama, murtuza.jadliwala\}@utsa.edu}
}

\maketitle

\begin{abstract}
We introduce \name, a novel DNS over HTTPS (DoH) protocol that leverages the InterPlanetary Name System (IPNS), along with public cloud infrastructure, to create a censorship-resistant moving target DoH service. \name~is specifically designed to evade traditional censorship methods that involve blocking DoH servers by IP addresses or domains by continually altering the server's network identifiers, significantly increasing the complexity of effectively censoring \name~traffic without disruption of other web traffic. We also present an analysis that quantifies the DNS query latency and financial costs of running our implementation of this protocol as a service. Further tests assess the ability of \name~to elude detection mechanisms, including both commercial firewall products and advanced machine learning-based detection systems. The results broadly support \name's efficacy as a robust, moving target DNS solution that can ensure continuous and secure internet access in environments with heavy DNS-based censorship.

\end{abstract}


\section{Introduction}
\label{sec:introduction}

The Domain Name System (DNS) is a crucial component of the Internet, responsible for translating human-readable domain names into machine-readable IP addresses. This service traditionally operates over port 53/udp and is integral to the functionality of the Web. However, traditional DNS queries are sent in plaintext, making them vulnerable to various attacks, including DNS spoofing, hijacking, and surveillance by third parties. These vulnerabilities are often exploited by governments and organizations that enforce censorship through DNS-based firewalls. DNS-based firewalls monitor DNS requests and can block access to specific domain names, thus restricting users from visiting certain websites or services. DNS-based firewalls have been used to enforce censorship against platforms such as Wikipedia, TikTok, and political websites in several regions~\cite{wikimedia_pakistan_2023,tiktok_compares_itself_2020,master2023modeling}.

To address these vulnerabilities, encrypted DNS protocols, such as DNS over TLS (DoT)~\cite{hu2016specification}, DNS over QUIC (DoQ)~\cite{huitema2022rfc}, DNSSEC~\cite{ateniese2001new}, and DNS over HTTPS (DoH)~\cite{hoffman2018dns}, were developed. %
DoT, DoQ, and DNSSEC encrypt DNS traffic, protecting it from eavesdropping, but operate on distinct ports (853/tcp, 853/udp, and 53/tcp, respectively), which make them easy to identify and block. In contrast, DoH integrates DNS requests with regular HTTPS traffic on port 443/tcp, the same port utilized for most encrypted web traffic. This makes it much harder for firewalls to block DoH without disrupting regular internet activities, as DoH is effectively masked within standard HTTPS traffic. As a result, DoH has the advantage of bypassing firewalls that block all outgoing DNS queries on port 53 or 853. %

Efforts to counteract DoH's ability to evade DNS-based censorship have led to several research initiatives focusing on specifically detecting and blocking DoH traffic. Early approaches relied primarily on list-based methods, which involve maintaining a static blocklist of well-known public DoH server IP addresses and domains to block them at the network-level firewall. However, this approach is limited due to the increasing ease of self-hosting DoH servers~\cite{adguard_home,pi_hole} on cloud infrastructure with vast IP address spaces that can make an infrequently updated blocklist ineffective. 
Entirely blocking outbound HTTPS traffic to hyperscalers such as AWS~\cite{aws}, Google Cloud~\cite{google_cloud}, or Azure~\cite{azure} is impractical for most censors as these hyperscalers also host numerous essential web services. As a result, blanket blocking of hyperscalers to target DoH services would cause significant unintended disruptions, making it an ineffective censorship strategy.
More sophisticated DoH detection mechanisms have also been developed that utilize machine learning (ML)-based techniques~\cite{jerabek2023dns,montazerishatoori2020detection,niktabe2024detection,mitsuhashi2022malicious}. These ML models analyze traffic flow patterns and other features to distinguish DoH traffic from regular HTTPS traffic. While some of these detection techniques have high false positive rates, their potential utility towards blocking DoH traffic emphasizes the necessity for more censorship-resistant DNS solutions.

In this paper, we introduce \name, a novel DoH client-server protocol designed to be censorship-resistant. The \name~protocol employs a moving target defense by dynamically changing the server IP address through the use of public cloud infrastructure, and securely sharing the latest server IP address with the client(s). 
\name's client-side software continuously updates the operating system to use the most recent server IP address for DoH queries. To mitigate propagation delays in sharing of new IP addresses with the client, the server temporarily keeps older IP addresses active (alongside the new IP address), ensuring continuous availability for clients in-between IP updates.
While there are various methods to securely share the latest server IP address, we leverage the InterPlanetary File System (IPFS)~\cite{ipfs} for its decentralized nature, which makes it more difficult for adversaries to detect or block~\cite{balduf2023m}. 
The \name~client integrates fully within the operating system, making it compatible with all browsers and applications without requiring any special configuration or additional plugin. This is in contrast to out-of-band DNS methods like DNS in Google Sheets~\cite{dns_google_sheets} or DNS over Discord~\cite{dns_over_discord}, which lack seamless integration with user environments.

This moving target architecture makes \name~highly resilient against list-based blocking methods, as the server IP addresses are frequently rotated, making it practically impossible to maintain an accurate blocklist. We also evaluate \name's effectiveness against ML-based detection methods, which attempt to analyze traffic patterns to block DoH queries. \name~demonstrates the ability to evade even these advanced detection systems while maintaining a high level of performance.

Our key contributions in this paper are as follows:
\begin{itemize}[leftmargin=*]
    \item \textbf{Design and Implementation of \name~Protocol:} We design and then implement the \name~protocol, a moving target DoH server that leverages public cloud infrastructure and IPNS to dynamically rotate its IP addresses, making it resilient against list-based DNS blocking and detection methods. %
    
    \item \textbf{Comprehensive Performance Evaluation:} We evaluate \name's performance in terms of DNS query latency and compare it against other DNS services, including well-known DoH providers and censorship-resistant alternatives like DoH over Tor. Our results demonstrate that \name~delivers low-latency performance, comparable to well-known public DoH services.
    
    \item \textbf{Evaluation of Censorship Resistance:} We empirically demonstrate \name's ability to evade both static, list-based blocking and more advanced ML-based detection systems that attempt to identify DoH traffic. This adaptability ensures the DoH service remains accessible in networks with heavy censorship.
    
    \item \textbf{Cost Analysis:} We provide analysis of \name's operational costs, showing that it is an affordable solution for individuals and organizations seeking a censorship-resistant DNS service. %
\end{itemize}

While \name~empowers users in censored networks to bypass DNS firewalls and freely access the internet, it simultaneously presents challenges for enterprise administrators who rely on protective DNS filtering for security~\cite{rodriguez2023two}. \name~can bypass such critical defenses, highlighting the dual nature of censorship-resistant tools. 

\section{Background and Related Work}
\label{sec:background-related}

\textbf{DNS.}
DNS is a fundamental component of the Internet, typically operating over port 53/udp to translate domain names into IP addresses. However, traditional DNS queries are in plaintext, leaving them vulnerable to attacks like DNS spoofing, hijacking, and surveillance~\cite{master2023modeling}. This lack of security also makes DNS an easy target for censorship by governments and organizations, who use DNS-based firewalls to block access to specific domains~\cite{wikimedia_pakistan_2023,tiktok_compares_itself_2020,master2023modeling,hoang2021great}.

\textbf{Encrypted DNS.}
DoT~\cite{hu2016specification}, DoQ~\cite{huitema2022rfc,kosek2022dns,kosek2022one}, and DNSSEC~\cite{ateniese2001new,lian2013measuring,chung2017longitudinal} were developed to encrypt DNS queries and enhance user privacy. %
DoT, DoQ, and DNSSEC protect DNS traffic from eavesdropping, but operate on distinct ports (853/tcp, 853/udp, and 53/tcp, respectively), making it easy for censors to detect and block~\cite{bottger2019empirical}.
When this traffic is blocked, users often have no choice but to use censoring DNS servers allowed in the censored network.

\textbf{DNS over HTTPS.} 
DoH encrypts DNS traffic using standard HTTPS protocol, which is typically transmitted over port 443/tcp~\cite{hoffman2018dns}. This allows DoH to blend in with regular web traffic, making it much harder for firewalls to detect and block without affecting normal HTTPS activity~\cite{bottger2019empirical}.
Further proposals to enhance the privacy of DoH have been introduced, including Oblivious DoH~\cite{singanamalla2021oblivious}.

\textbf{DNS Blocking and Filtering.}
Several previous studies have examined the extent of DNS-based censorship~\cite{master2023modeling,hoang2021great,pearce2017global,wander2014measurement,farnan2019analysing,hoang2019measuring,niaki2020iclab,hoang2024gfweb,bhaskar2022many,burnett2013making,kuhrer2015going,calle2024toward,bock2020detecting,sundara2020censored}, wherein networks of various scales block access to external (unfiltered) DNS servers and mandate the use of DNS servers that provide manipulated or filtered responses.
Jin et al.~\cite{jin2021understanding} found that using encrypted DNS resolvers allowed access to 37\% of censored domains from vantage points in China, whereas none of the censored domains were accessible from Iran due to additional censorship methods such as SNI-based blocking of the websites~\cite{shbair2016improving,bock2021even}.
SNI-based blocking of websites can be bypassed using other censorship circumvention tools~\cite{shbair2015efficiently,satija2021blindtls}.

\textbf{DoH Detection and Blocking.}
Efforts to block DoH traffic have evolved from static blocklists of known DoH server IPs and domains to more advanced detection mechanisms utilizing ML techniques~\cite{jerabek2023dns,montazerishatoori2020detection,niktabe2024detection,mitsuhashi2022malicious}. These models analyze traffic flows and other features to distinguish DoH traffic from regular HTTPS. Although some of these techniques suffer from high false-positive rates, this growing capability to block DoH traffic highlights the need for more resilient solutions like \name.%

\textbf{DNS Censorship Circumvention.}
DNS over Tor routes DNS queries through the Tor network to anonymize both the user and their DNS traffic~\cite{dns_tor}. Unlike DoT, DoQ, DNSSEC, and DoH, which focus only on encrypting the communication between the client and resolver, DNS over Tor is focused on protecting user anonymity by masking the origin (and the destination) of the queries. However, the onion routing used by Tor can result in significantly increased latency for DNS queries, which may lead to slower browsing experiences.
There are also alternative out-of-band DNS methods, such as DNS over Google Sheets~\cite{dns_google_sheets} and DNS over Discord~\cite{dns_over_discord}, though these approaches lack the seamless integration with user environments.

\textbf{Internet Censorship Circumvention.}
Censorship-circumvention tools such as Tor~\cite{dingledine2004tor} and VPNs~\cite{scott1999virtual,feilner2006openvpn,donenfeld2017wireguard} enable users to bypass restrictions imposed by censored networks. 
Importantly, both tools ensure that DNS queries also travel through their respective networks, allowing users to bypass local blocks on the use of external (uncensored) DNS servers. 
However, both Tor and VPNs can themselves be blocked through techniques such as stateful deep packet inspection, IP filtering, or by identifying and blocking the traffic patterns associated with their protocols~\cite{winter2012great,singh2017characterizing,xue2022openvpn,zain2019vpn}.
There have also been efforts to bypass these blocking techniques~\cite{wang2017your,dixon2016network,green_tunnel}, but this remains an ongoing challenge, as advancements in evasion strategies are continually met with increasingly sophisticated censorship mechanisms.

\textbf{Moving Target Defense.} Moving target defense strategies strengthen system resilience by continuously modifying system characteristics, adding complexity that disrupts adversaries’ ability to predict and exploit vulnerabilities~\cite{jajodia2011moving, zhuang2014towards, cho2020toward}. It is widely applied across domains, including enhancing software security~\cite{williams2016shuffler, larsen2014sok, jackson2011compiler, baudry2015multiple, snow2013just, jang2016breaking, seo2017sgx}, establishing communication channels resistant to interference~\cite{gao2018game, quan2015anti, zhang2023rl}, protecting virtual machines hosted on shared infrastructure~\cite{torquato2021analysis, torquato2025evaluation}, defending against DDoS attacks~\cite{aydeger2024mtdns, wright2016moving, zhou2021toward, zhang2022mitigate, nguyen2022moving}, and securing critical infrastructure~\cite{heydari2016securing, jafarian2012openflow, casola2022designing, luo2014effectiveness, xu2021adaptive}.
Similar in concept to our proposed system, Domain Generation Algorithms (DGAs) have been used as a moving target defense by malicious actors, dynamically creating new domain names to sustain communication with command and control (C2) centers and evade detection~\cite{sood2016taxonomy,fu2017stealthy,nie2023pkdga}. However, DGAs are unsuitable for \name's censorship-resistant goals because they can be susceptible to detection algorithms without frequent domain and IP changes~\cite{li2019domain, anderson2016deepdga}, and are constrained by ICANN's one-year minimum domain registration requirement, making frequent domain changes costly.

\section{Adversary Model}
\label{sec:adversary}

\begin{figure}[t]
    \centering
    \includegraphics[width=0.85\linewidth]{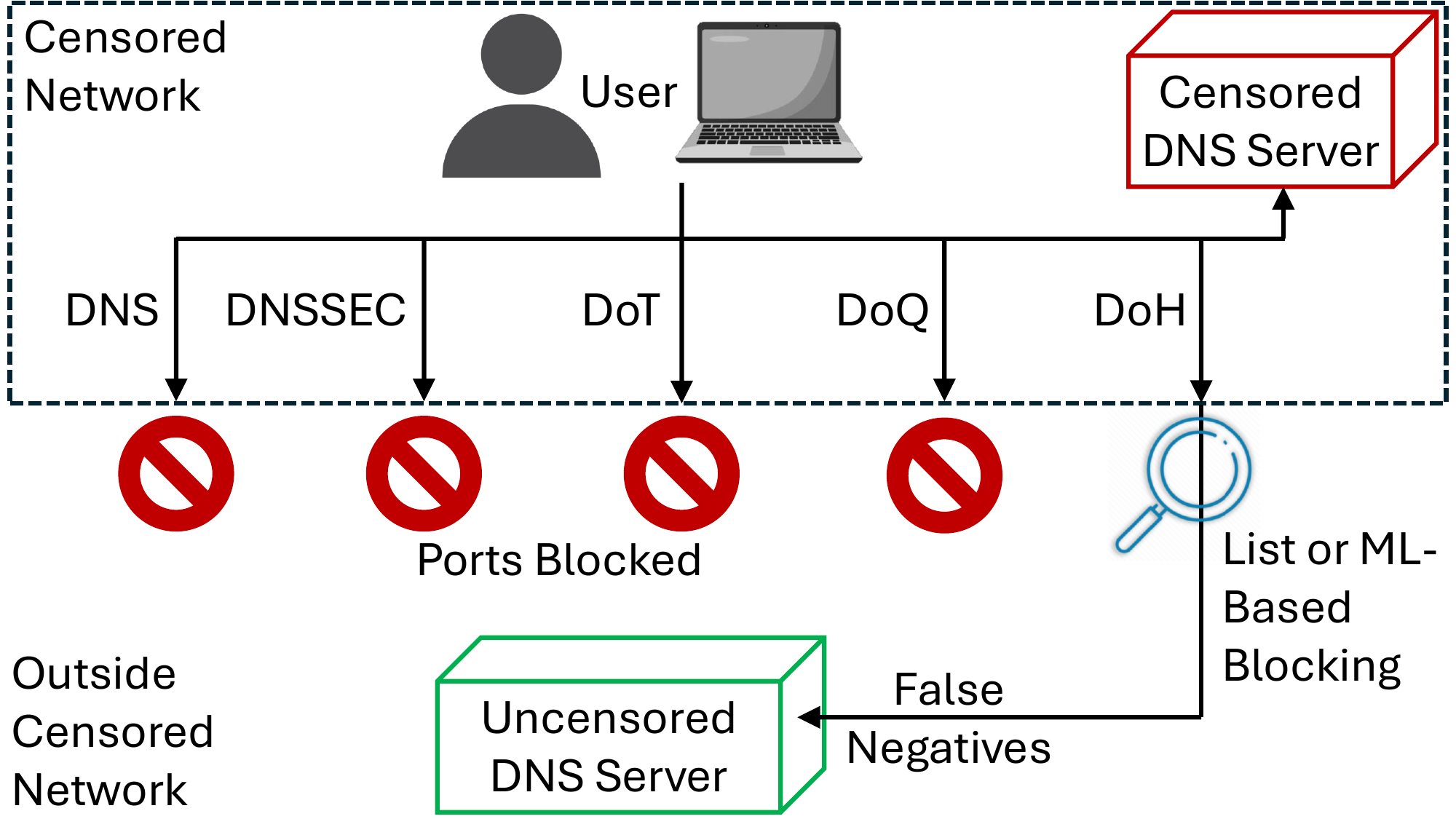}
    \caption{Overview of the adversary model.}
    \label{fig:adversary}
\end{figure}

Our adversary model for \name~assumes that censors implement DNS-based censorship by blocking access to external DNS resolvers and forcing users to rely on their own DNS servers with filtering rules (\cref{fig:adversary}). The censor controls the network, which may be managed by an ISP, an enterprise network administrator, or a government authority with regulatory oversight. %
We categorize the adversary's capabilities into two following groups:

\begin{itemize}[leftmargin=*]
    \item \textbf{Blocking by IP or Domain Name}: Censors may attempt to block DoH traffic by maintaining blocklists of IP addresses or domains associated with DoH servers. \name~aims to counter this by employing dynamic IP address rotation using hyperscalers, making it difficult to maintain effective blocklists.

    \item \textbf{Machine Learning-Based Detection}: Advanced adversaries may use ML techniques to identify DoH traffic based on traffic flow characteristics. Although state-of-the-art ML models are capable of detecting regular DoH traffic, \name~aims to evade detection by minimizing flow durations by employing IP address rotation and using randomized DoH query paths to defeat active probing.

\end{itemize}

\name~is effective under the following assumptions about the censorship environment:

\begin{itemize}[leftmargin=*]
    \item \textbf{DNS-based Censorship}: The primary method of censorship is blocking external DNS resolvers and forcing users to use the censors' DNS servers that filter specific websites. \name~aims to provide a way to bypass these DNS restrictions. %

    \item \textbf{Websites' IP Addresses Not Blocked}: It is assumed that the IP addresses of the websites themselves are not blocked at the network-level firewall. If censors block entire IP ranges, \name~may not provide access to those websites unless additional bypass techniques~\cite{wang2017your,dixon2016network,green_tunnel,shbair2015efficiently,satija2021blindtls} are employed.

    \item \textbf{No Endpoint Control}: The adversary does not have control over the client endpoints. This implies that the adversary cannot force SSL/TLS interception or proxy use with SSL/TLS decryption, and users are free to install any software on their devices (such as the \name~client). %

    \item \textbf{Trusted Cloud Provider}: \name~requires a trusted hyperscaler to host its server instance, ensuring that the cloud provider does not collude with the adversary to identify instances running \name~server code. %
    
    \item \textbf{Use with VPNs or Tor}: \name~is not required on a VPN or the Tor network, as both of these tools can already enable access to external DNS resolvers. But if \name~is used in parallel with VPNs or Tor, it may provide low-latency DNS resolution, avoiding the higher latency caused by routing DNS queries through these networks. Moreover, in environments where VPNs and Tor are blocked, leaving users without access to uncensored DNS, \name~aims to provide an effective solution.

    \item \textbf{Private Client-Server Communication}: \name~is designed to be hosted by individuals or small groups, ensuring that the secret shared between the server and the clients, which is used to communicate IP address changes, remains private. This prevents the adversary from knowing the latest IP addresses and subsequently blocking them.

\end{itemize}

\section{\name~System Model}
\label{sec:system}

\begin{figure*}[ht]
    \centering
    \includegraphics[width=\linewidth]{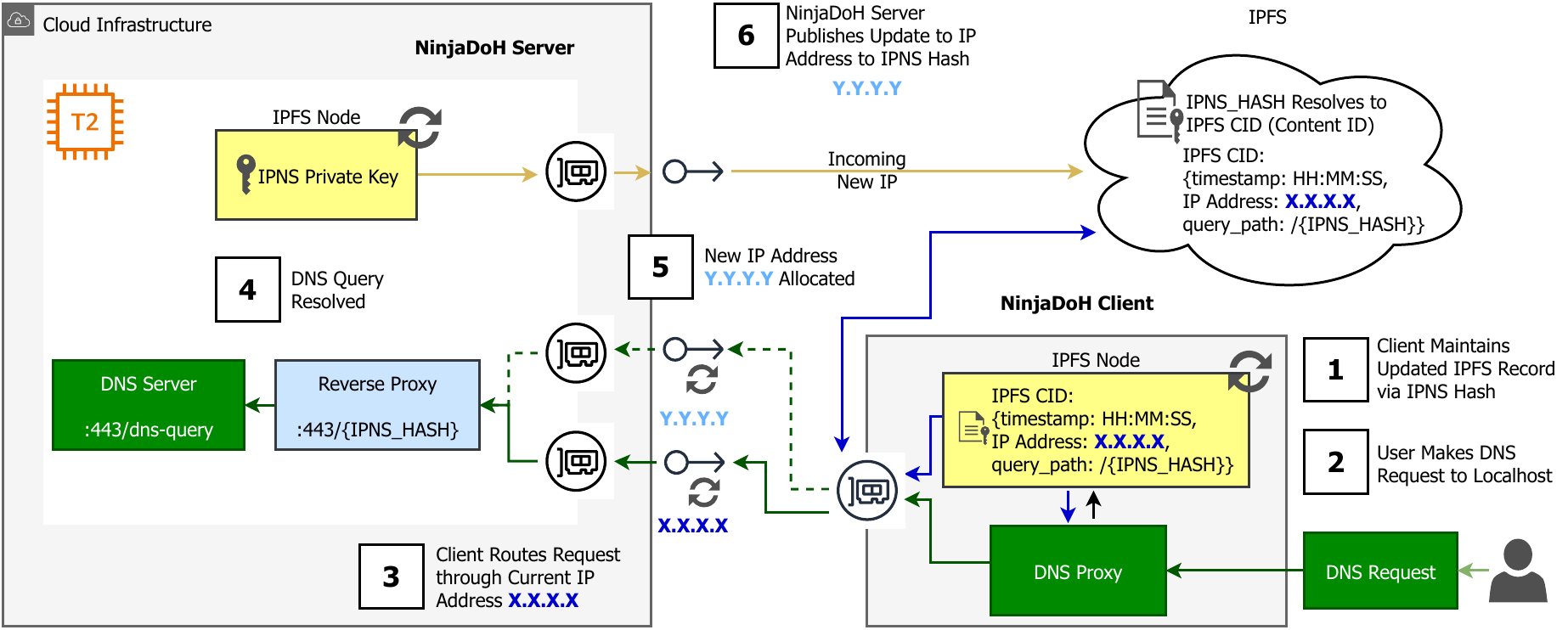}
    \caption{Overview of the \name~protocol and architecture. The \name~client maintains an updated IPNS record with the latest server information via IPFS. When the user initiates a DNS request, it is first sent to the client's localhost DNS proxy, which then routes the request to the current IP address of the \name~server. At a configurable frequency, the server allocates a new IP address and publishes this update to IPNS. The client retrieves the updated IP address via IPNS and uses it for future DNS queries (until next IP update).}%
    \label{fig:system}
\end{figure*}

\subsection{System Requirements}
\label{subsec:system-reqs}

We define six requirements that address key dimensions of performance, resilience, scalability, and usability. \textbf{R1:} The system should provide low-latency DoH services, ensuring it performs efficiently under realistic workloads. \textbf{R2:} The system should outperform or match the performance of DoH over Tor, offering a practical and competitive alternative for censorship resistance. \textbf{R3:} It should bypass firewalls that rely on blocklists of known domains, IP addresses, or patterns, maintaining uninterrupted traffic flow in censored environments. \textbf{R4:} The system must evade detection by machine learning models, which includes three sub-requirements: \textbf{R4a:} It should resist detection by baseline ML detection models, ensuring resilience against existing state-of-the-art methods for identifying DoH traffic; \textbf{R4b:} It should remain resistant to adaptive adversaries, preventing detection even when adversaries specifically train their models on the system's traffic; and \textbf{R4c:} It must introduce scalability challenges for adversarial ML-based detection, making it impractical for adversaries to apply detection models at large scale. \textbf{R5:} The system should be cost-effective, enabling deployment and operation with reasonable resource consumption. \textbf{R6:} Finally, the system must afford a seamless and user-friendly experience, ensuring minimal disruption to regular internet usage and compatibility with common user environments.
To meet these requirements, we present the \name~client-server architecture as shown in~\cref{fig:system}.

\subsection{The Server}

\textbf{IP Address Space and Networking.} The cornerstone of \name~is that it is a moving target DoH service. What makes it a moving target is that its IP address changes frequently, at a configurable frequency, a capability only made possible by having a large IP address allocation pool available to quickly swap in and swap out. Therefore, a public cloud provider that has access to such an IP allocation pool is a key component of this system design. We selected Amazon Web Services (AWS) as our public cloud provider, but we believe that any public cloud provider that allows for the dynamic allocation of IP addresses to a compute instance would be a suitable infrastructure layer for the system. At time of writing, the AWS IPv4 public address pool is estimated to be over 100 million addresses.

For the prototype system, three elastic network interfaces (ENI) were created inside a single region virtual private cloud (VPC). Each ENI functions as a virtual network interface card that can be dynamically assigned an IP address from AWS's IPv4 address pool. One of the ENIs serves to separate management plane traffic from the application traffic. The other two ENIs serve as the primary and alternate interfaces for application plane traffic. Both of these two application plane ENIs will route traffic to the same reverse proxy listening on all available interfaces. The \name~protocol can scale to an arbitrary number of ENIs in a deployment, given the compute instance can support the desired number of ENIs.

\textbf{Compute.} The system only requires modest compute resources, as the server runs a few lightweight services and does not require much persistent storage. The system was tested on a \textit{t2.small} instance, which has 1vCPU and 2GB of RAM. Additionally, Ubuntu server's small operating system footprint and short interval persistent log keeping means that the block storage requirement is about 20GB. 

\textbf{Software.} Architecturally, the system requires a reverse proxy and a recursive DNS on the server. The reverse proxy has to be capable of forwarding HTTPS traffic from an arbitrary path from the application ENIs to the DNS service. The recursive DNS software has to be able to resolve DoH requests. Both have to be able to accept custom certificates and modify their certificates on-the-fly or with a short reload (so as not to affect the user experience). There are several software packages capable of meeting these requirements, but for our prototype we chose two widely deployed, free and open source software packages. The key external software dependencies of our implementation are \textit{nginx}~\cite{nginx} and \textit{AdGuard Home}~\cite{adguard_home}. A key, useful feature of \textit{AdGuard Home} is that it can be configured to use multiple upstream DoH providers to resolve the client's query. When configured with several reputable upstream providers, there is a gained benefit of the user not creating an identifiable signature on any specific upstream provider. 

\textbf{Certificate Management.} Certificate management ensures secure communication between clients and the DoH service. Given that \name~frequently rotates its public IP addresses, maintaining valid SSL/TLS certificates for these changing IP addresses presents unique challenges. Traditional certificate issuance processes are not designed for such dynamic environments, necessitating a custom solution. To address this, \name~employs an automated certificate generation and management process that dynamically creates and signs certificates whenever the set of public IP addresses changes. This process leverages a private Certificate Authority (CA) hosted on the server, allowing for rapid certificate issuance without relying on external CAs. Clients are configured to trust this private CA on setup, allowing them to trust new certificates on-the-fly despite the frequent IP changes.

The certificate management workflow is integrated into the system's orchestration code and operates as follows:

\begin{enumerate}[leftmargin=*]
    \item \textbf{Detection of IP Address Changes}: When the system allocates a new IP and associates it with one of the application ENIs, it triggers the certificate regeneration routine. This ensures that the certificates always reflect the current set of public IP addresses.

    \item \textbf{Certificate Signing Request (CSR) Creation}: A CSR is created using the private key. The Common Name (CN) in the CSR is set to the newest IP address in the current list. An extension file is generated to include all current public IP addresses under the \texttt{subjectAltName} field. This ensures that the certificate is valid for any of the IP addresses clients may connect to.

    \item \textbf{Certificate Signing \& Reload}: The CSR is signed using the private CA's key and certificate to produce the server certificate. The certificate includes the SANs specified. After the new certificate is generated, the reverse proxy is reloaded to the updated certificate, ensuring that incoming connections are secured using the latest certificate without significant downtime.

\end{enumerate}

\textbf{Manipulating the Query Path.} A DoH query requires a query path to reach the DoH service on the server. While not specifically required by the DoH RFC, Bottger et al.~\cite{bottger2019empirical} show that the majority of DoH providers use the query path cited in the RFC examples, \texttt{/dns-query}. This  pseudo-standard has been exploited as a technique to actively probe if an address is hosting a DoH service~\cite{jerabek2023dns}. Therefore, the \name~protocol calls for using a non-standard query path. For our prototype, we set the query path as the IPNS hash. We propose that for a production environment, this path be randomized with varying lengths as to further obfuscate its packet structure signature. 

By making this custom query path the only accessible path for accessing the DoH service, \name~prevents potential DoH downgrade attacks~\cite{huang2020comprehensive} and defeats identification via active probing on other common DNS server ports (such as 53 and 853).

\textbf{IPFS Node.} The InterPlanetary File System (IPFS) node distributes the updated public IP address to clients in a decentralized and censorship-resistant manner. Recent work has shown that IPFS maintains its functionality even under some of the most oppressive censorship regimes~\cite{balduf2023m}. By leveraging IPFS and the InterPlanetary Name System (IPNS), \name~disseminates routing information, without relying on traditional DNS infrastructure or via a censorable storage or communication medium. This process ensures that clients always have access to the current IP address, even as it changes frequently due to the moving target defense strategy.

\begin{algorithm}[t]
\caption{NinjaDoHServerRoutine()}\label{alg:server_routine} \small
\KwData{Compute instance with set of network interfaces $l$; global address pool $p$; subset of IPs $p' \subset p$ where one IP maps to one interface $f(p' \rightarrow l$)}
\KwResult{IP address rotated and published via IPNS}

\While{Service is running}{
    \textbf{Define:} $l_x$ as the interface with the oldest IP
    
    \Comment{Add new IP from $p$ into $p'$ and assign it to $l_x$}
    newIPAddress $\gets$ PullNewIP($p$)\;
    AssignIPToInterface(newIPAddress, $l_x$)\;

    \Comment{Release disassociated IP back to the pool $p$}
    ReleaseDisassociatedIPs($p'$)\;
    
    \Comment{Use $p'$ to update certs}
    UpdateSSLCertificates($p'$)\;

    \Comment{Generate a random query path}
    queryPath $\gets$ GenerateRandomQueryPath()\;

    \Comment{Publish $p'$ via IPNS}
    content $\gets$ CreateContent($p'$, queryPath, getTimestamp())\;
    cid $\gets$ PublishToIPFS(content)\;
    UpdateIPNSRecord(cid)\;

    \Comment{Wait for next scheduled rotation}
    Sleep(rotationInterval)\;
}
\end{algorithm}

The workflow for updating and publishing the IP address via IPFS and IPNS is integrated into the system's orchestration code and operates as follows:

\begin{enumerate}[leftmargin=*]    
    \item \textbf{Adding Content to IPFS}: Upon allocating a new Elastic IP (EIP), the system creates a JSON object that includes the new IP address, a query path, and a timestamp. The JSON object is added to the local IPFS node using the IPFS HTTP API. The system receives a Content Identifier (CID) for the added content, which uniquely references the data in the IPFS network. %

    \item \textbf{Publishing to IPNS}: The system publishes the CID to IPNS using its private key. %
    IPNS allows for mutable pointers to immutable content in IPFS, enabling clients to resolve the latest CID associated with a given IPNS address. The \texttt{PublishToIPFS} and \texttt{UpdateIPNSRecord} functions in~\cref{alg:server_routine} make the new CID available to the client via IPNS using the specified key name.

    \item \textbf{Verification and Propagation}: After publishing, the system verifies the new IPNS record by attempting to resolve it locally and may publish the new CID to the IPFS PubSub system to expedite propagation to connected peers. IPNS PubSub is experimental and we show in the evaluation that normal resolution via the DHT is sufficient. 

\end{enumerate}

The most critical step in this process is publishing the new CID to IPNS, which ensures that clients can always retrieve the latest IP address using the stable IPNS address. The parameters in the implementation code ensure that the publication is forced, resolves any conflicts, and sets appropriate lifetimes and time-to-live (TTL) values. After publishing, the function verifies the IPNS record by resolving it. By utilizing IPFS and IPNS, we avoid dependence on centralized services, which can be manipulated or blocked by adversaries. 

This approach ensures continuous service availability by creating a ``ladder'' of IP addresses. As the system rotates IP addresses, clients connected via older IPs can still maintain their connections because each network interface always has a valid IP address assigned. By configuring a sufficient number of network interfaces and selecting an appropriate rotation interval, the system guarantees overlap between the old and new IP addresses. This overlap allows clients enough time to update to the new IP address before the one they are using is released back to the pool. Unlike a load balancer, which distributes traffic across multiple servers, \name~rotates IP addresses on a single server without redirecting client requests. This strategy ensures there is no downtime for valid IP addresses, preventing dropped requests and maintaining a seamless user experience during IP rotations.

\subsection{The Client}

The client component of \name~is designed to securely and reliably connect to the moving target DoH service provided by the server. Given the frequent rotation of the server's IP addresses and the utilization of a private CA, the client must dynamically update its configuration to maintain seamless connectivity. To achieve this, the client leverages its own access to IPFS to retrieve the latest server information.

\textbf{IPNS Key Exchange \& Root Certificate Installation.} A foundational aspect of the client's operation is the secure exchange of a secret between the client and server during the initial setup. This secret is the IPNS name hash used by the server to publish its current IP address in IPFS. By possessing this IPNS key, the client can resolve the server's latest IP address. This mechanism ensures that only authorized clients, who have obtained the IPNS key through a secure channel, can access the service.

Securely exchanging this secret presents a challenge, especially in environments where communication channels may be monitored or censored. To address this, we propose several mechanisms for clandestinely and securely conducting the secret exchange, drawing on prior work in secure communication and key distribution. %
One approach is to utilize out-of-band channels that are less likely to be monitored or censored. Steganographic techniques can embed the secret within innocuous-looking media files or messages. Prior research has demonstrated the effectiveness of steganography for covert communication in restrictive environments~\cite{kadhim2019comprehensive}. By embedding the IPNS key within images, audio files, or other media, the secret can be transmitted without arousing suspicion. Another mechanism is to leverage encrypted messaging applications that provide end-to-end encryption and forward secrecy, such as Signal or WhatsApp~\cite{unger2015sok,cohn2020formal,rosler2018more}. These platforms are designed to resist surveillance and interception, making them suitable for securely transmitting the IPNS key and root certificate~\cite{johansen2021snowden}.

Since the server uses a private CA to issue SSL/TLS certificates for its frequently changing IP addresses, clients must additionally install the server's root CA certificate on their machines. The transmission of the private CA certificate can be accomplished in a similar manner as the IPNS hash, or once the IPNS hash has been secretly received, the client can pull the private CA and the client software from IPFS.

\begin{algorithm}[t]
\caption{NinjaDoHClientRoutine()}\label{alg:client_routine} \small
\KwData{IPNS key $k$; IPFS API endpoint $u$; DNS proxy configuration path $c$}
\KwResult{Continuous connectivity to the DoH server with dynamic IP updates}

\While{Client is running}{
    \Comment{Resolve the latest IPNS content}
    cid $\gets$ ResolveIPNS($k$, $u$)\;

    \If{cid is valid}{
        \Comment{Retrieve IP address and query path from IPFS}
        (ip, queryPath) $\gets$ GetIPAndQueryPathFromIPFS(cid, $u$)\;

        \If{ip is new}{
            \Comment{Update the DNS proxy config}
            UpdateDNSConfig(ip, queryPath, $c$)\;
            RelaodDNSProxy($c$)\;

            \Comment{Log new IP address}
            UpdateLastKnownIP(ip)\;
        }

        \Comment{Test DoH connectivity}
        connectivity $\gets$ TestDoHConnectivity(ip)\;
        UpdateClientStatus(connectivity)\;
    }
    
    \Comment{Sleep before next update check}
    Sleep(updateInterval)\;
}
\end{algorithm}

\textbf{Client Software and Operation.} We implement a prototype \name~client in Python, as a terminal application that performs the key functions to ensure continuous access to a \name~server as seen in the~\cref{fig:screenshot} screenshot. It interacts with a local IPFS node to resolve the IPNS key and retrieve the latest CID with the server's IP address information. Using the CID, the client fetches the JSON object containing the server's current IP address, query path, and timestamp. If the retrieved IP address differs from the one previously used, the client updates its local DNS resolver configuration accordingly. The client could use any other means to retrieve IPFS content, such as an IPFS gateway or access to a non-local IPFS node. 

To update the DNS resolver configuration, the client utilizes \texttt{dnscrypt-proxy}\footnote{https://github.com/DNSCrypt/dnscrypt-proxy}, a flexible DNS proxy that supports encrypted DNS protocols, including DoH. The client generates a new DNS stamp based on the latest IP address and query path, then modifies the \texttt{dnscrypt-proxy} configuration file to include this updated stamp. Following the configuration update, the client reloads the proxy to apply the changes, ensuring that subsequent DNS queries are forwarded to the updated DoH server address.

Moreover, our implementation of the client continuously monitors connectivity to the server by performing DNS queries and measuring latency. %
Our interface offers users immediate feedback about the system's status and any connectivity problems. Additionally, \textit{without a valid connection to the \name~server, DNS queries will fail, which will prevent unintentional DNS leaks.} \name~is resistant to DoH downgrade attacks~\cite{huang2020comprehensive} as long as browsers and applications are configured to use the operating system's default DNS settings, which are continuously updated by the \name~client.

To ensure timely updates, the client can listen for changes on the IPNS PubSub topic associated with the shared IPNS key or at some configurable frequency (for our implementation every 5 seconds), to resolve updates via IPNS. When the server publishes a new update, indicating a change in the service's IP address, the client retrieves the latest information and repeats the configuration update process. This mechanism allows the client to follow the server's moving target defense strategy without manual intervention of the end user. As long as the client is able to update the IP address of the server before that IP address is rotated out completely (a function of how many ENIs the server uses and its rotation interval), the client will retain service. One of the added benefits of a \name~client is that it maintains compatibility with existing operating systems, software, and the current DNS system, unlike out-of-band methods, as discussed previously. 

\textbf{Dependencies.} The client relies on a few key dependencies and environmental requirements. An IPFS node or gateway must be accessible to resolve IPNS records and retrieve content from the IPFS network. A capable DNS proxy must be installed to handle encrypted DNS queries and support dynamic updates of the server's stamp. The client routine is illustrated in~\cref{alg:client_routine}.

\section{Evaluation}
\label{sec:evaluation}

We designed the system to meet the system requirements in \cref{subsec:system-reqs}, informed by the adversary model in \cref{sec:adversary}. Now, using our implementation of the system, we evaluate it against the system requirements.

\subsection{Baselining Latency}

To baseline the latency between the different DoH providers, we queried the top domains from the Cisco Umbrella Top 1M list\footnote{https://umbrella-static.s3-us-west-1.amazonaws.com/index.html}, recording both ping and query response times. We used randomized subdomains to help prevent upstream caching.

\textbf{Public DoH Providers.} We evaluated the performance of different DNS resolvers, including the \name~server, a Control %
server that was hosted on the same AWS region with just \textit{AdGuard Home} configured identically to the \name~implementation, and a group of Public DNS resolvers (Cloudflare, Google, and Hurricane Electric (HE)). The primary metric is the DNS resolution time, which is the ping adjusted time to return the DNS query measured in milliseconds (\SI{}{\milli\second}).

\cref{tab:mean_dns_times} presents the mean DNS resolution times for each server type. The \name~server demonstrated an average resolution time of \textbf{\SI{12.68}{\milli\second}}, comparable to the Control server's resolution time of \textbf{\SI{7.85}{\milli\second}}, indicating negligible performance differences. The Public DNS group demonstrated an average resolution time of \textbf{\SI{7.77}{\milli\second}}.

\begin{table}[h]
    \centering \scriptsize
    \caption{Mean ping-adjusted DNS resolution times by DoH server type.}
    \label{tab:mean_dns_times}
    \begin{tabular}{|l|c|c|}
        \hline
        \textbf{Server Type} & \textbf{Mean Resolution Time (\SI{}{\milli\second})} & \textbf{Confidence Interval (95\%)} \\
        \hline
        \name~& 12.68, n=4000 & (11.54, 13.81) \\
        Control & 7.85, n=4000 & (7.38, 8.32) \\
        Public DNS & 7.77, n=12000 & (7.23, 8.31) \\
        \hline
    \end{tabular}
\end{table}

The confidence intervals indicate that the difference between the \name~and both the Control server and public DoH providers is, for an end user, practically negligible, where our system adds about \qtyrange[range-units=single,range-phrase=--]{4}{5}{\milli\second} of latency to each request. This can be explained by both the added step of the client making a call to its local proxy instead of directly to the server and that a slight delay during IP rotation. During this collection period the \name~rotated IP addresses three times and had zero dropped client queries. 

\textbf{DNS over Tor.} In this section, we compare the performance of \name~with DNS over Tor. The primary metric of comparison remains the DNS resolution time. DNS over Tor provides censorship-resistance by routing DNS requests through the Tor network, but it comes at the cost of increased latency. For the DNS over Tor experiments, we used Cloudflare's DoH service on Tor\footnote{dns4torpnlfs2ifuz2s2yf3fc7rdmsbhm6rw75euj35pac6ap25zgqad.onion}. In total, 10 Tor circuits were used to collect data, with each circuit performing DoH lookups for 10 domains. The experiments were conducted by repeatedly generating new Tor circuits using the \texttt{NEWNYM} signal to obtain fresh exit nodes.

\begin{table}[h]
    \centering
    \caption{DNS resolution time comparison (ping-adjusted): \name~vs. DNS over Tor.}
    \label{tab:tor_vs_ninjadoh}
    \begin{tabular}{|l|c|c|c|c}
        \hline
        \textbf{Server} & \textbf{Mean (\SI{}{\milli\second})} & \textbf{Median (\SI{}{\milli\second})} & \textbf{Std Dev (\SI{}{\milli\second})} \\
        \hline
        \name~& 12.68 & 4.31 & 23.77 \\
        DNS over Tor & 601.16 & 562.00 & 376.00 \\
        \hline
    \end{tabular}
\end{table}

\begin{figure}[t]
    \centering
    \includegraphics[width=0.9\linewidth]{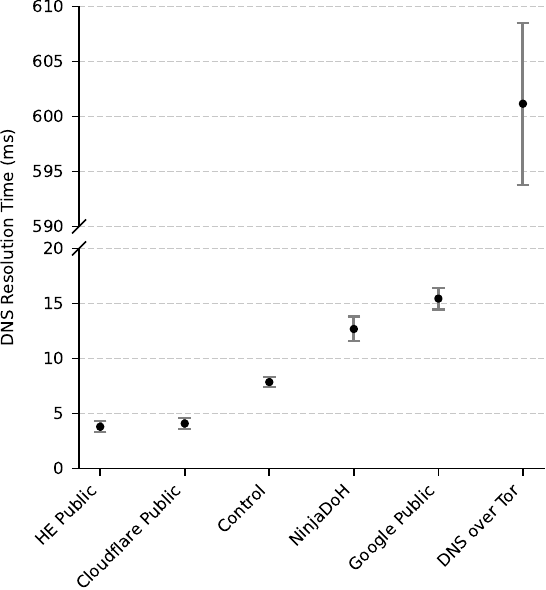}
    \caption{Mean DNS resolution time with 95\% confidence intervals for different DNS servers. The $y$-axis includes a break to show the large discrepancy between standard DNS servers and DNS over Tor.}
    \label{fig:dns_latency_comparison}
\end{figure}

\cref{tab:tor_vs_ninjadoh} presents the summary of DNS resolution time statistics for \name~and DNS over Tor. The \name~setup provided a significantly lower average resolution time, with a mean ping adjusted query time of \textbf{\SI{12.68}{\milli\second}}, compared to DNS over Tor, which exhibited a much higher mean resolution time of \textbf{\SI{601.16}{\milli\second}}. This large discrepancy in DNS resolution times highlights the latency overhead introduced by Tor's network routing. \cref{fig:dns_latency_comparison} shows the combined, ping adjusted mean query resolution times for the public DoH providers, the control, NinjaDoH, and DNS over Tor.

\begin{summarybox}{Fast and Efficient DoH Service (R1 \& R2)}
    Our evaluation confirms that \name~provides low-latency DoH services with an average resolution time of \SI{12.68}{\milli\second}. This is only about \SI{4.91}{\milli\second} slower than public DNS resolvers (average \SI{7.77}{\milli\second}). In stark contrast, DNS over Tor exhibits a significantly higher average resolution time of \SI{601.16}{\milli\second}, demonstrating the latency overhead of Tor's network routing. These results fulfill R1 by demonstrating efficient performance under realistic workloads and R2 by offering a practical alternative to existing censorship-resistant techniques with minimal additional latency.
\end{summarybox}

\subsection{List-based Firewall Evasion}

List-based firewall blocking is the most prevalent method to restrict access to DoH services and enforce censorship policies. These firewalls maintain static blocklists of known DoH server domains and IP addresses, often sourced from public repositories. These can be curated lists specifically targeting DoH providers, or in the most extreme, lists to block all known public DNS providers.\footnote{https://github.com/curl/curl/wiki/DNS-over-HTTPS}\footnote{https://public-dns.info/nameservers.txt} Commercial providers advertise capabilities that they have the most up-to-date and relevant blocklists, and some governments and ISPs offer a Protective DNS (PDNS) service to their constituents and users which rely on such updated lists~\cite{rodriguez2023two}. By blocking these known entities, censors aim to compel users to either fallback to unencrypted DNS queries, or use resolvers that they control, which can then be readily surveilled. 

To evaluate the efficacy of \name~in evading such list-based blocking methods, we evaluated it against techniques across OSI layers. As summarized in \cref{tab:firewall_evasion}, \name~successfully bypasses domain and IP blocking by not utilizing a domain name and by dynamically changing its IP address. Additionally, it evades application-level identification by mimicking regular HTTPS traffic, avoiding typical DoH traffic patterns through mechanisms like changing the query path that could trigger deep packet inspection filters. While \name~cannot circumvent strict IP or domain allowlisting, where all unknown IPs and domains are blocked, this method is highly restrictive, costly to maintain, and impractical for most network environments.

\renewcommand{\tabularxcolumn}[1]{m{#1}}

\begin{table*}[h]
    \centering
    \caption{Evasion of DoH blocking by popular firewalls.}
    \label{tab:firewall_evasion}
    \begin{tabularx}{\textwidth}{
        |>{\centering\arraybackslash}m{2.5cm}
        |>{\centering\arraybackslash}m{2cm}
        |>{\centering\arraybackslash}m{1.2cm}
        |>{\raggedright\arraybackslash}m{5cm}
        |>{\raggedright\arraybackslash}X| }
        \hline
        \textbf{DoH Blocking Method} & \textbf{Layer} & \textbf{Evaded} & \textbf{Adversarial Impact} & \textbf{Explanation} \\
        \hline
        Domain Blocking & Network (3) & \ding{51} & Low-cost to implement and maintain as it relies on simple static blocklists. Minimal impact on the network and does not significantly degrade user experience. & Does not use a domain name so bypasses static domain blocklists. \\
        \hline
        IP Blocking & Network (3) & \ding{51} & Low-cost to implement and maintain as it relies on simple static blocklists. Minimal impact on the network and does not significantly degrade user experience. & The IP address agility of the system evades static IP blocklist. \\
        \hline
        Application Identification & Application (7) & \ding{51} & Moderate cost and complexity as it requires deep packet inspection. It can introduce slight latency but is more effective than simple IP or domain blocking. & Mimics regular HTTPS traffic to evade application-level filters. Does not follow the typical patterns of DoH traffic. \\
        \hline
        SNI Blocking & Session (5) & \ding{51} & Moderate cost with little impact on user experience. More complex to maintain with newer protocols like TLS 1.3, which makes SNI blocking harder. & Since the client connects via an IP address and not a hostname, there is no \texttt{ClientHello} message sent. \\
        \hline
        IP Allowlisting & Network (3) & \ding{55} & High cost and impact, as it restricts user access significantly and requires substantial maintenance of trusted IP addresses. This method limits user capabilities and heavily controls the network environment. & Uses strict IP allowlisting; unknown IPs are blocked, so dynamic IPs are ineffective. \\
        \hline
    \end{tabularx}
\end{table*}

\begin{summarybox}{Bypasses Firewall Blocklists (R3)}
    \name~successfully evades 4 out of 5 common DoH blocking methods employed by popular firewalls, including domain blocking, IP blocking, application identification, and SNI blocking. By not utilizing domain names, dynamically changing IP addresses, and mimicking regular HTTPS traffic, \name~maintains uninterrupted traffic flow in censored environments where blocklists are employed. This fulfills R3 by effectively bypassing current, prevalent censorship techniques.
\end{summarybox}

\subsection{Machine Learning Evasion}
\label{subsec:evaluation}

To evaluate the performance of \name~against machine learning (ML) detection techniques, various models from previous works have been selected. ML-based inspection of HTTPS traffic has become increasingly prevalent, and DoH is not inherently immune. Using the methodology of Jerabek et al.~\cite{jerabek2023dns} with tooling from MontazeriShatoori et al.~\cite{montazerishatoori2020detection} (hereafter referred to as DoHlyzer\footnote{https://github.com/ahlashkari/DoHLyzer}), five state-of-the-art model architectures were evaluated. 

\textbf{Models}. Four of the models were Deep Neural Networks (DNNs). The \textit{LSTM-Based Model} uses Long Short-Term Memory units to capture temporal dependencies in network traffic sequences, making it suitable for time-series data. \textit{The Fully Dense Model} processes traffic flows independently, which allows for faster computation but may miss sequential patterns. The \textit{CNN-Based Model} employs 1D Convolutional Neural Networks to detect local patterns within the data, identifying spatial or temporal correlations. Finally the \textit{Hybrid LSTM-Dense Model} combines dense and LSTM layers, balancing feature extraction with sequential modeling for improved performance. Each DNN model was trained using different flow sequence lengths (from 4 to 10) to assess how traffic sequence length impacts detection performance. This allows for evaluation of both short-term and long-term traffic patterns. For the decision tree classifier, we reproduce the XGBoost model by Jerabek et al.~\cite{jerabek2023dns} with original code and training data. XGBoost is a gradient boosting decision tree algorithm known for its effectiveness in structured data analysis. 

\textbf{Datasets}. The training dataset used for the DNN models was sourced from the DoHLyzer repository, containing both DoH and non-DoH samples with time-series features. For the XGBoost model, the training dataset consists of PCAP files extracted from the \texttt{DoH-Gen-F-AABBC} database~\cite{jevrabek2022collection}. This dataset features encrypted traffic enriched with key TLS attributes such as \texttt{TLS\_ALPN}, \texttt{TLS\_JA3}, and \texttt{TLS\_SNI}, which provide rich feature sets for training and evaluating ML models. Additionally, the dataset includes a list of known, public DoH providers' IP addresses.

\textbf{Flow Stitching.} To prepare the training data, the traffic was consolidated into bidirectional flows, grouping packets based on common characteristics, such as source and destination IP addresses, ports, protocols, and timestamps, creating a comprehensive view of the entire traffic session. Some of the extracted features of the bidirectional flows from the PCAP files are \textit{Mean Payload Size, Number of Packets, Client-to-Server Packet Ratio, and Mean Time Between Packets.} A flow is classified as DoH or non-DoH by the IP address list of known DoH IP addresses found in the dataset.

\begin{figure}[b]
    \centering
    \includegraphics[width=\linewidth]{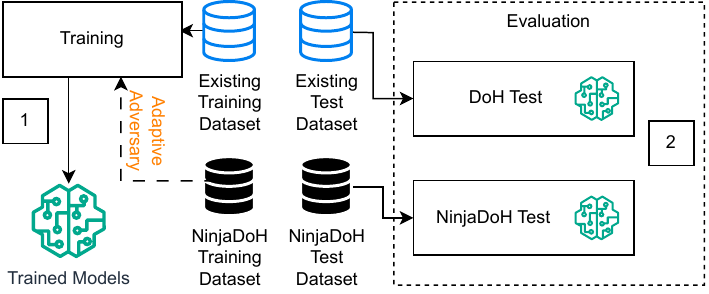}
    \caption{Evaluation model to detect \name~traffic.}%
    \label{fig:ml-training-eval}
\end{figure}

\textbf{Training.} The dataset was split it into a training set and a test set. The models were then trained on the training set (Step 1 in \cref{fig:ml-training-eval}). Following the training phase, the models were evaluated on the test set (Step 2 in \cref{fig:ml-training-eval}). The results from this evaluation on the baseline training data is consistent with the target from the prior work. F1 scores, precision, and recall metrics all ranging between 0.98 and 0.99, demonstrating the effectiveness of the models in detecting DoH traffic.
Each DNN model was trained with varying flow sequence lengths, ranging from 4 to 10 timesteps. The sequence length, or the number of timesteps, plays a crucial role in capturing temporal dependencies within the data, especially for models like LSTMs that are designed for sequential information. By experimenting with different sequence lengths, we aimed to find the optimal configuration that maximized the model's performance. After training and evaluating the models across the various sequence lengths, the best sequence number for each model was determined based on the highest F1 score achieved during testing to balance  precision and recall.

\begin{figure}[t]
    \centering
    \includegraphics[width=\linewidth]{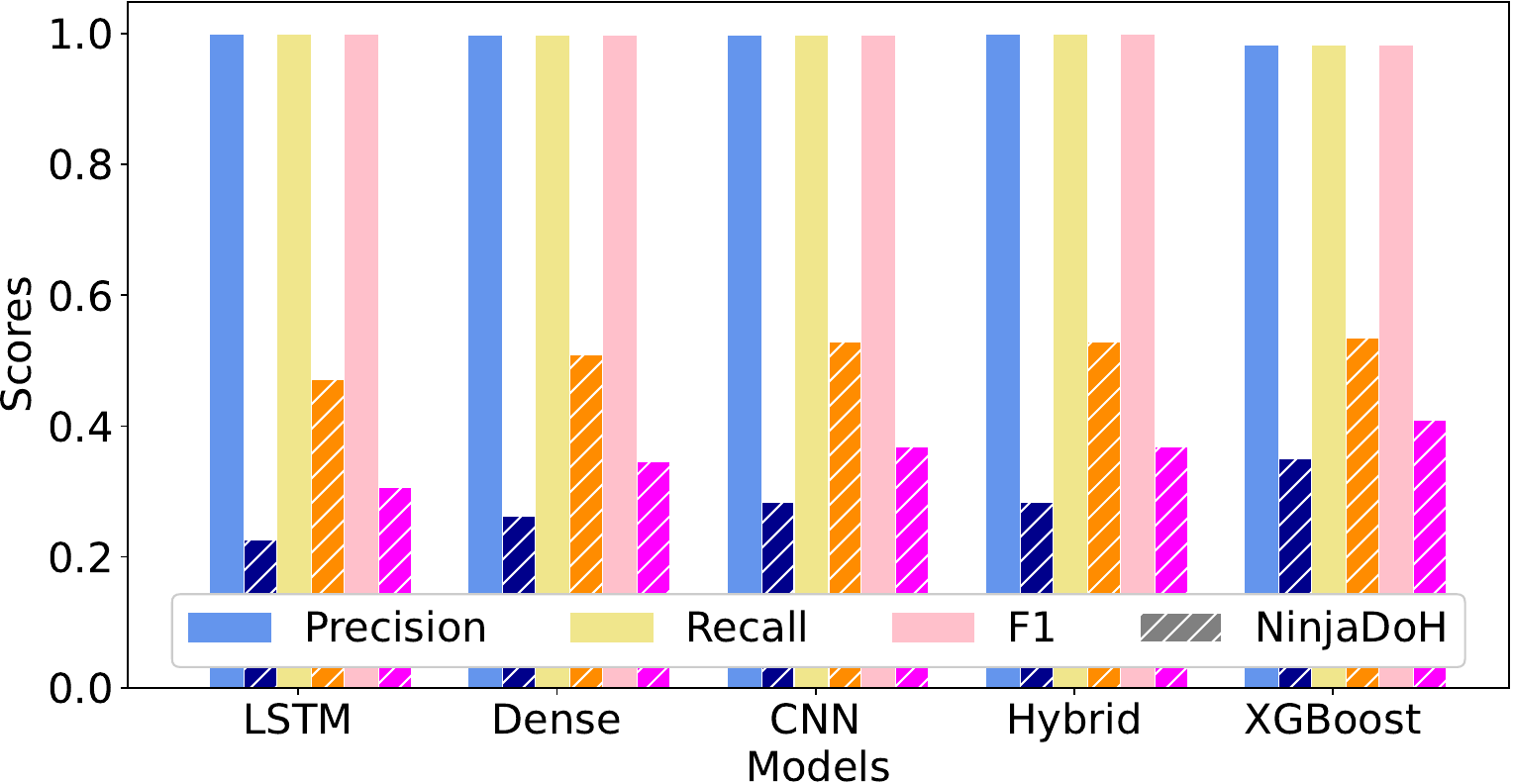}
    \caption{Performance of ML models trained with DoH traffic in detecting test DoH vs. \name~traffic, showing that while regular DoH is accurately detected, \name's dynamic IP rotation and query path randomization make detection far more challenging.}
    \label{fig:ml-baseline}
\end{figure}

\textbf{Results.} We then evaluated the trained models against new traffic, this time captured from a \name~session. We captured three-minute's worth of typical browsing activity by an end-user who was using \name, which became the evaluation set. The resulting PCAP file is processed to bidirectional flow format for compatibility with the models. Each trained model is subsequently given the evaluation set to see how well it can detect \name~traffic. In this scenario, the adversary faces trade-offs: false positives impact network usability, emphasizing the importance of precision, while missed detections (recall) reduce the adversary's effectiveness in censorship.
  
\begin{summarybox}{Undetectable by Current ML Models (R4a)}
    The system effectively resists detection by baseline ML detection models. As shown in the \cref{fig:ml-baseline}, the models achieve an average recall of 0.506, which is comparable to random guessing. Precision and F1-scores similarly reflect poor detection capability. This meets R4a by demonstrating resilience against existing state-of-the-art methods for identifying DoH traffic, rendering the detection models ineffective against \name~traffic.
\end{summarybox}

\begin{figure}[b]
    \centering
    \includegraphics[width=\linewidth]{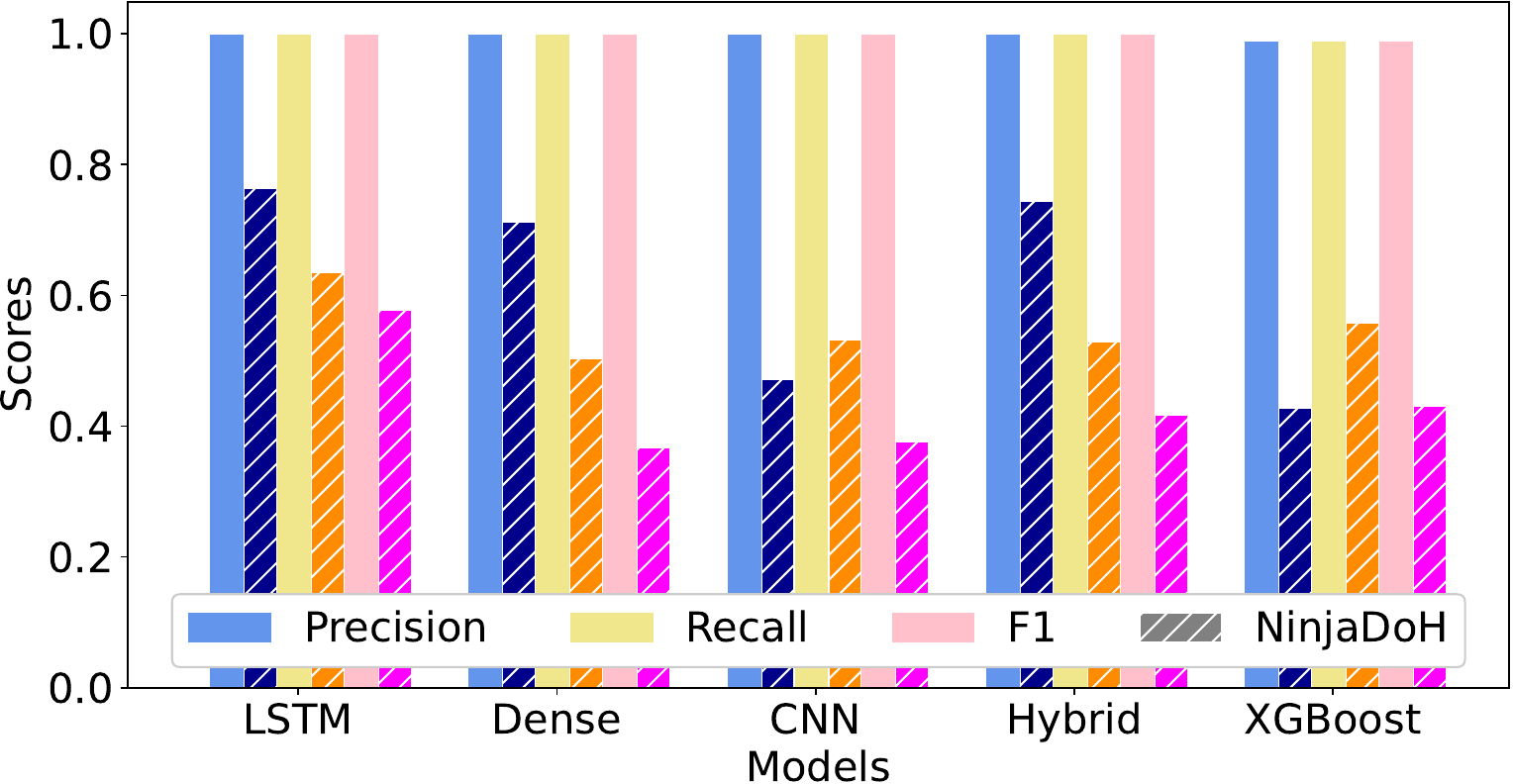}
    \caption{Performance of ML models trained with DoH and \name~traffic in detecting test DoH vs. \name~traffic.}
    \label{fig:ml-adaptive}
\end{figure}

\subsection{Adaptive Adversary Evasion}

In our evaluations so far, we have assumed a static adversary who employs their current censorship strategies and is either unaware of, or specifically not targeting \name~users. In this section, we evaluate our system against a notional, adaptive adversary that is both aware of, and specifically targeting \name. This adversary can deploy their own \name~server outside their network, and then monitor the traffic from a client that they control within their network. There are several configurable parameters to a \name~deployment that an attacker wouldn't know, such as the IP rotation interval (the interval itself could be stochastic), the IP address blocks in use, or any additional traffic shaping techniques employed by the system deployer. However, they would know the general architecture and be able to deploy their own instance and experiment with their own parameters

Niktabe et al.~\cite{niktabe2024detection} show that linear models specifically trained on known malicious DoH traffic data can have performance similar to deep neural networks in classification performance. This illustrates the potential power of an adversary who specifically trains a model against known \name~traffic. Linear models will have a faster inference time than deep neural networks in evaluating new traffic. However, the authors note that there still exists ``Inadequate detection of obfuscated and disguised malicious traffic.'' Therefore, even if a model is specifically trained on \name~traffic, it will have to be able to conduct its inference fast enough in order to actually deny the service at a practical level.

Most previously known DoH providers are eventually censored by these approaches because once the censor has determined a domain or IP is classified as a DoH provider, that domain or IP is added to a blocklist. Since \name~is IP-agile and doesn't use domain names, \textit{even if an adaptive adversary has a perfect model for detecting \name~traffic, they will only be able to temporarily deny service after the user's first query until the next IP rotation.} 

To evaluate this, we assumed the role of a determined adversary that deployed their own system with perfect visibility into ground-truth network traffic. The system was configured with a pseudo-random IP rotation frequency between \qtyrange[range-units=single,range-phrase=--]{1}{3}{\text{minutes}}. Data was collected over a \SI{15}{\text{minute}} period, which resulted in 10 distinct IP rotations and a $>$\SI{1}{\giga\byte} PCAP file to use for training. The traffic was captured and processed in the same manner as in \cref{subsec:evaluation}. This ground-truth \name~traffic is mixed with both ``benign'' DoH traffic and non-DoH traffic for training the models. The newly trained adversarial model was used on the same evaluation set as before, and \cref{fig:ml-adaptive} shows that performance improves across all the model types that we evaluated with the \name~traffic present in the training set.

There are a number of reasons why this tailored model shows improvement. In the data processing step, flows are extracted from the \texttt{pcap} file, and then are processed into both statistical and time series features. Both of these feature sets that the models are trained on use flow sequences of packets between the DoH server and the client. We found that the \name~traffic produced a normally-distributed number of ``clumps'' of packets in each flow sequence, with $\mu=10.30$, $\sigma=5.20$. However, the benign DoH traffic produced much higher average clump sizes ($\mu=135.64$) and a reflects a log-normal distribution with $\sigma=1.57$, $ \lambda=26.68$. Therefore, it possible that the sequences of \name~traffic from one IP to the next are features that can be learned, leading to the model gaining predictive power from the information-sparse \name~flows. However, this improved performance does not currently approach an acceptable level for a would-be censor. 

\begin{summarybox}{Resilience to Adaptive Adversaries (R4b)}
    Even when facing adaptive adversaries who specifically train models on \name~traffic, detection performance improves but remains insufficient for effective censorship. The best model trained with \name~traffic achieves a precision of 0.764, recall of 0.635, and F1-score of 0.578. While these metrics are higher than those without targeted training, they do not reach levels that would allow a censor to reliably detect and block \name~traffic. This fulfills R4b by showing that the system remains resistant to adversaries who tailor their models to detect its traffic.
\end{summarybox}

\subsection{Scalability of Adversarial ML-based Detection}

Detecting DoH traffic in large networks using ML models presents significant scalability challenges, especially against \name. To block traffic effectively, an adversary must capture enough packets to reconstruct flows, extract features, and perform inference before the \name~client rotates to a new IP address. As network scale increases, the feasibility of this approach diminishes due to the growing data volume and computational demands.

To illustrate the computational burden on an adversary, we model the detection process and estimate the time required to process network traffic at different scales. Our model simulates the adversary's detection process, incorporating components that reflect real-world network conditions and constraints.

Flow arrivals are modeled as a Poisson process to represent the random nature of flow initiations in a network. The inter-arrival times ($T_a$) are exponentially distributed with rate $\lambda$, where $\lambda = \frac{\text{Flows per Minute}}{60}$. Flow durations ($D$) are modeled using a log-normal distribution to reflect the positive skew observed in real data, as evidenced by our collected flow duration statistics (\cref{tab:flow_duration}). This distribution accounts for the wide variability in flow lengths, with many short flows and a few long-lasting ones.

\begin{table}[t]
    \centering \small
    \caption{Descriptive statistics for flow duration (in \SI{}{\milli\second}).}
    \label{tab:flow_duration}
    \begin{tabular}{|l|c|c|c|c|}
        \hline
        \textbf{Flow Type} & \textbf{Mean} & \textbf{Median} & \textbf{Skew} & \textbf{N} \\
        \hline
        Non-DoH  & 7430 & 313 & 3820 & 3395 \\
        \name    & 742  & 246 & 4240 & 1606 \\
        Combined & 5280 & 265 & 4740 & 5001 \\
        \hline
    \end{tabular}
\end{table}

\begin{figure}[t]
    \centering
    \includegraphics[width=\linewidth]{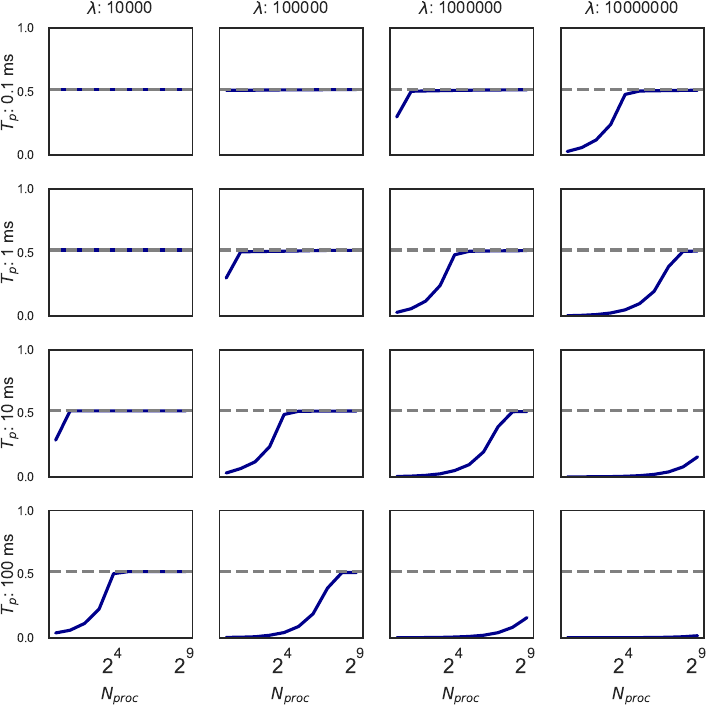}
    \caption{Relationship between the number of cores ($N_{\text{proc}}$), flows/min ($\lambda$), and flow processing time ($T_p$) on the probability of detecting DoH. The dashed line represents the best ML model's true positive rate baseline (TPR=0.52).}
    \label{fig:flow_size}
\end{figure}

The adversary must process each flow within the IP rotation interval ($T_{\text{rotation}} = \SI{60}{\second}$)
, with a fixed processing time per flow ($T_p$), such as \SI{0.1}{\milli\second}, \SI{1}{\milli\second}, or \SI{10}{\milli\second}. Processing starts after the flow ends and a processor becomes available, calculated as $S_i = \max(E_i, A_j)$, where $E_i$ is the flow end time and $A_j$ is the processor's availability.

The performance metric of interest is the probability of detecting DoH flows ($P_{\text{detect\_DoH}}$). This is calculated by dividing the expected number of detected DoH flows by the total number of DoH flows. The expected detected DoH flows are computed as the product of the processed DoH flows and the true positive rate (TPR), i.e., $F_{\text{processed\_DoH}} \times \text{TPR}$.

\[
P_{\text{detect\_DoH}} = \min\left( 1,\ \dfrac{N_{\text{proc}}}{\lambda\, T_p} \right) \times \text{TPR}
\]

\cref{fig:flow_size} presents the simulation results that show that unless the adversary can process all flows in near real-time, the probability of detecting DoH traffic declines significantly. Achieving this requires substantial computational resources, which is impractical and cost-prohibitive at scale. This scaling becomes even more impractical when considering the additional overhead of packet capture, flow reconstruction, and feature extraction required for the detection process.

Moreover, even if the adversary manages to detect and block an IP address, the benefit is short-lived due to frequent IP rotations. The adversary would need continuous detection process, incurring ongoing computational costs. This imposes a disproportionate burden on the adversary compared to potential minimal disruption caused to \name~users. This asymmetry underscores the inherent scalability challenges faced by adversaries attempting to detect and block DoH traffic in large-scale networks.

\begin{summarybox}{Scaling ML Detection is Impractical (R4c)}
    Our modeling suggests that adversaries face significant scalability challenges when attempting to detect \name~traffic using ML-based methods at large scale. Even with such resources, the detection probability remains limited by the ML model's true positive rate. This satisfies R4c by demonstrating that widespread ML-based detection is infeasible due to the disproportionate computational costs involved.
\end{summarybox}

\subsection{Costs to Deploy \& Integrate} 

At time of writing, to implement our prototype of the server-side, the price is \$23.55 USD/month. Even cheaper infrastructure for the server may be available on different cloud providers, or by using more lightweight DNS resolver software.  Monthly subscriptions for VPNs for a single user cost between \$10 and \$15, so this is a cost efficient system that can deliver censorship resistant DoH-as-a-service to many users for slightly more than a single VPN subscription. While this is not a VPN, it is useful for comparison in terms of cost to an end user. As seen in \cref{fig:screenshot} in Appendix~\ref{appendix:screenshot}, the screenshot of the prototype, the client implementation integrates into the end-user's existing operating system and workflow with no additional configuration or software to configure.

\begin{summarybox}{Affordable and User-Friendly (R5 \& R6)}
    The system fulfills R5 by offering cost-efficiency for individual users, while scaling up to larger cloud instances remains affordable for privacy-focused groups and communities. It satisfies R6 through OS-level client integration, ensuring seamless compatibility with existing operating systems, software, and the current DNS infrastructure. 
\end{summarybox}

\section{Additional Security Analysis}
\label{sec:sec-analysis}

Now that we have evaluated the empirical performance of \name~against the most common types of DoH blocking methods, including an adaptive adversary, we analyze what additional security considerations are relevant for our proposed system. 

\textbf{IPv4 Reuse and Short-term Tenancy.} Although the global IPv4 address pool of hyperscalers is incredibly large, by constantly rotating IPv4 addresses there exists the possibility that IP addresses are recycled and reused. There could be unintended consequences of using transitory IP addresses, and service could be denied if an in-use IP address was previously blocked by a network administrator because of the prior tenant of that address. IP reuse on public clouds is an area of active research, as Pauley et al. show~\cite{pauley2023dscope,pauley2022measuring}, cloud IPs receive significantly more targeting scanning than what is expected under normal conditions. This can be caused by misconfigurations of prior tenants, or that cloud IPs appear as more lucrative targets. This should have minimal affect on our system because of the transitory nature of our tenancy on any given IP address. 

Let's assume an adversary gains access to the IPNS hash allowing them to know the IP addresses and follow along with the IP rotations of a particular \name~deployment. This would allow them to add those IP addresses to their blocklist. While this would deny use of that deployment, it would cause downstream negative effects on legitimate services that become tenants of those IP addresses. This negative effect on the censor would be compounded as the number of individual \name~deployments they block by using a compromised IPNS hash. This furthers the case against append-only, list-based approaches to blocking DoH. Additionally, to restore service, the \name~deployer would rotate their IPNS hash and securely communicate the hash rotation with their clients. 

\textbf{IPv6 Subnet Blocking.} As IPv6 is orders of magnitude larger an address space than IPv4, with no real risk of exhaustion, list-based approaches for IP addresses blocking are not possible. However, that doesn't mean that censors can't take advantage of certain administrative realities present in how IPv6 subnets are allocated to users. The smallest allocation to an individual entity is $/64$. Therefore, if a censor detects \name~on IPv6, they could choose to block the entire $/64$ subnet associated with that IP. This would largely come down a specific deployment consideration on how the public cloud provider will allocate IPv6 subnets to a user account. 

\textbf{Trusted Execution Environments (TEE).} A sophisticated adversary who gains root access to the server instance providing the service would have the ability to either deny service by terminating the process, or most insidiously, conduct a DNS poisoning attack by hijacking the DNS resolver located on the instance. If the infrastructure serving the \name~service is compromised, we want to limit the adversary to, at most, a denial of service. We can maintain the integrity of the system and preserve the privacy of the DNS requests of the client by placing key cryptographic functions of the system inside a trusted execution environment. The large hyperscale public cloud providers allow for secure compute enclaves on their compute instances, which Demigha et al.~\cite{demigha2021hardware} show can provide secure storage as well as protection from an OS-level compromise. The TEE in the cloud environment can hold the private key for the TLS session, as well as the private key for the IPNS name. Properly implementing this would ensure that even if the adversary gains root access on the host compute instance, they would not be able to manipulate the code, or access the private keys stored inside the TEE, maintaining the integrity of the service.

\section{Discussion and Future Work}
\label{sec:discussion}

We introduced the \name~protocol, a censorship-resistant DoH service that leverages the programmability and scale of hyperscalers with decentralized P2P networks to evade detection and blocking. Our evaluation demonstrates that \name~significantly outperforms existing censorship-resistant DoH techniques in terms of latency, while maintaining robust resistance against both list-based and machine learning-based censorship methods.

\textbf{Performance and Usability.} Our results indicate that \name~provides a solution that is 50 times more performant in terms of latency compared to the leading censorship-resistant DoH technique that utilizes the Tor network. Specifically, \name~achieves low-latency DNS query resolution comparable to standard DoH services, whereas DoH over Tor suffers from substantial latency overhead due to the nature of the Tor network's routing mechanisms.
Furthermore, we offer users the flexibility to deploy their own infrastructure or use a trusted public deployment. This contrasts with relying on the Tor network's relays and exit nodes, which can introduce additional vulnerabilities and are beyond the user's control\cite{chao2024systematic}. By eliminating dependence on the Tor network, we reduce potential attack surfaces and improves overall security.

\textbf{Novel Use of Hyperscalers.} The advent of cloud computing and decentralized networks (web3) is transforming how end-users interact with the internet. Traditional application architecture paradigms decoupled infrastructure from the application logic. However, now that infrastructure can be software-defined, they can become a key component of the application itself. \name~capitalizes on this shift by employing cloud computing resources to dynamically rotate IP addresses, making it difficult for adversaries to maintain effective blocklists. 
Simultaneously, the use of P2P overlay networks like IPFS allows for decentralized distribution of service information without reliance on traditional DNS infrastructure, which can be censored or manipulated. This novel combination of cloud agility and P2P networking to address an application-level security problem in DNS is a significant contribution of our work. It highlights how not only software solutions but also innovative use of on-demand, software-defined infrastructure can address complex security challenges.

\textbf{Evaluation of Censorship Resistance.} Our evaluation shows that the censorship-resistant features of \name~do not introduce practical overhead in latency or DNS query resolution performance. The system's deployment costs are modest and it operates without dependencies on proprietary software, making it accessible and cost-effective.

Importantly, we demonstrate robust resistance against current methods of blocking DoH traffic. It effectively evades list-based blocking techniques due to its dynamic IP rotation and lack of reliance on domain names. Additionally, our experiments indicate that \name~is protected against existing machine learning-based detection methods that utilize statistical or time-series based feature sets to train detection models. Even when an adaptive adversary specifically trains models on \name~traffic, the cost to implement a real-time system with acceptable false positive rates is high, suggesting that it can withstand sophisticated censorship attempts.

\textbf{Limitations and Future Work.} While we address several challenges in providing censorship-resistant DNS services, there are areas for future exploration. One limitation is that highly restrictive environments employing strict IP whitelisting or controlled client endpoints can still impede \name's operation. Future work could investigate integrating techniques to further obscure detectable signatures of the system, such as traffic mimicking or randomization strategies~\cite{tschantz2016sok}.

Moreover, expanding upon the techniques presented, future research could explore leveraging the scale and infrastructure agility provided by public cloud providers to tackle other security challenges. By harnessing software-defined, on-demand infrastructure, new avenues can be opened for developing resilient services in the face of evolving adversarial tactics.

\textbf{Ethical Considerations.}
Following Tor Research Safety Board guidelines~\cite{tor_safety_board}, Tor was used solely to measure DNS over Tor latency, without altering operations or impacting users. No Tor user data was collected or analyzed, ensuring data minimization and privacy.
Censorship-resistant DNS services like \name~is essential for individuals seeking to bypass oppressive state censorship and freely access information. However, this capability may introduce security concerns for managed networks that depend on protective DNS filtering to guard against malware and restrict harmful content. While state-level DNS blocking can serve as a tool for public control, similar methods are often necessary for network administrators to enforce security and compliance policies. Thus, while technologies that evade censorship promote open access, they can also weaken security measures in organizational contexts, requiring a careful balance between accessibility and protection needs.

\section*{Conclusion}
\label{sec:conclusion}

We introduced \name, a novel DoH protocol that uses IPNS and cloud infrastructure to create a censorship-resistant, moving-target DNS service. \name~is designed to bypass censorship techniques that block DNS servers by IP or domain by continuously changing network identifiers, making censorship of \name~traffic challenging without affecting other web traffic. We quantified DNS query latency and service cost, and evaluated \name's resilience against detection by commercial firewalls and ML-based systems. The results confirmed \name's effectiveness as a robust DNS solution, providing continuous, secure access in heavily censored environments.

\balance
\bibliographystyle{IEEEtran}
\bibliography{references}

\appendices

\clearpage

\onecolumn
\section{}
\label{appendix:screenshot}

\begin{figure}[H]
    \centering
    \includegraphics[width=\linewidth]{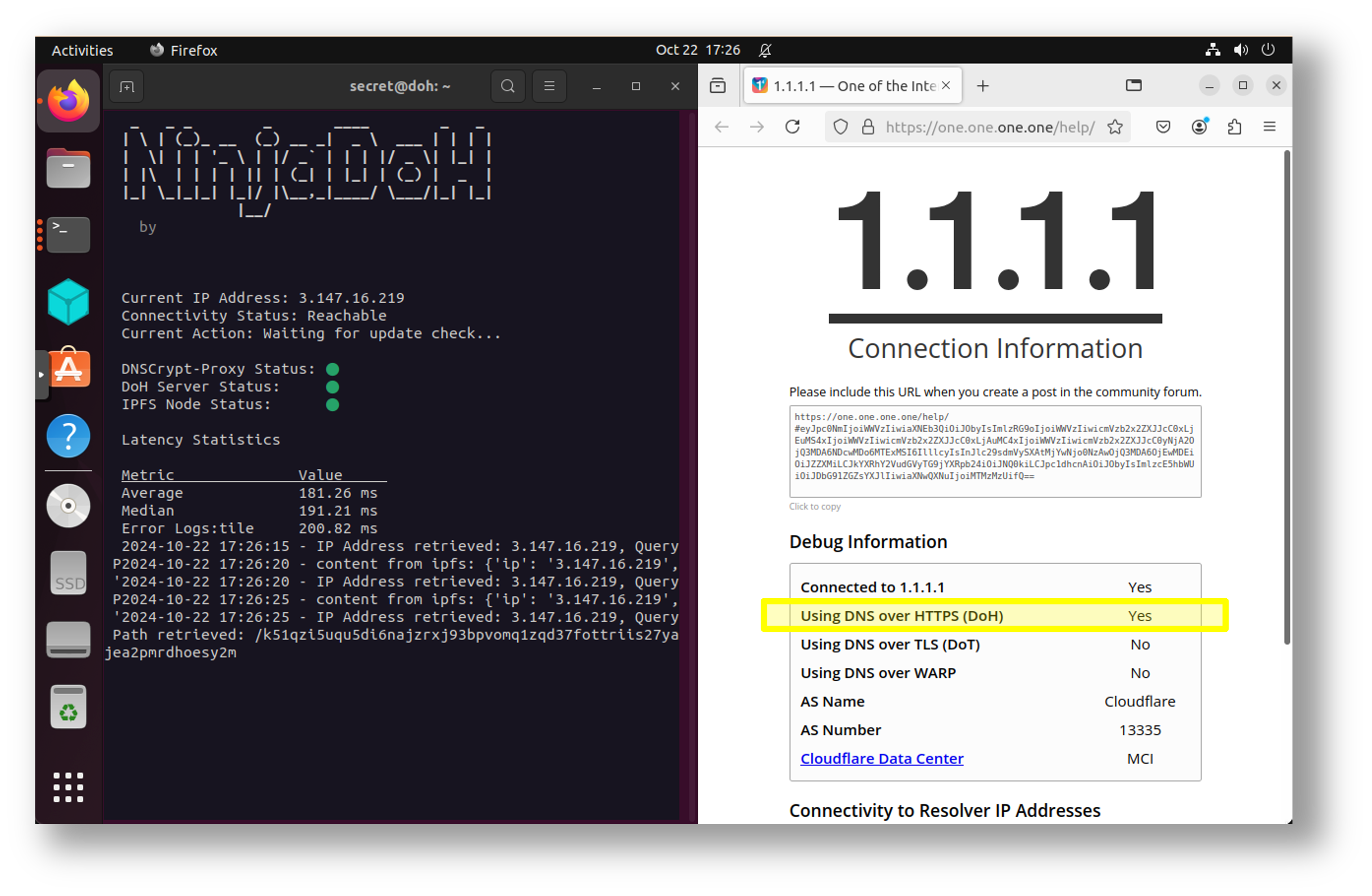}
    \caption{Prototype \name~client (left window) running on Ubuntu Desktop, using a \name~server instance on AWS for DNS queries. DoH connectivity to \name~server is verified by \url{https://one.one.one.one/help/} opened on a web browser (right window). Cloudflare is configured as the upstream DNS server on this instance of \name~server, as verified in the highlighted box.}
    \label{fig:screenshot}
\end{figure}

\end{document}